\documentclass[twoside]{dis08}
\usepackage[latin1]{inputenc}
\usepackage[dvips]{graphicx,epsfig,color}
\usepackage{wrapfig,rotating}
\usepackage{amssymb,amsmath,array}

\begin{document}
\title{Searches for Non-Standard-Model Higgs Bosons at the Tevatron}

\author{Chris Hays
%\thanks{}
\vspace{.3cm}\\
University of Oxford, Department of Physics \\
Oxford OX1 3RH, United Kingdom 
}

\maketitle

The Standard Model of particle physics has a number of asymmetries suggestive of 
a more fundamental theory, where the symmetries are restored.  If the symmetry 
breaking scale is the electroweak scale, new Higgs bosons beyond that of the 
Standard Model could be observed at the Tevatron.  Recent Tevatron searches have 
set the most stringent limits on new Higgs bosons in a variety of models.

\section{Introduction}
\label{sec:intro}
\hspace*{0.18in}
In the Standard Model (SM), the only symmetry broken by the Higgs mechanism is 
the electroweak symmetry.  However, the SM possesses unexplained features 
suggestive of a more fundamental theory, with additional symmetries broken at a 
mass scale that is as yet beyond our reach.  Two examples are the symmetry 
between fermions and bosons (supersymmetry) and between left- and right-handed 
fermion couplings to the weak gauge bosons (left-right symmetry).  These 
models predict additional Higgs bosons that could be observable at the 
Tevatron.  Recent Tevatron searches have probed for neutral Higgs bosons 
predicted by supersymmetry, and for doubly charged Higgs bosons in left-right 
symmetric models.

\section{Higgs Boson Searches in Supersymmetric Models}
\label{sec:susyhiggs}
\hspace*{0.18in}
A self-consistent supersymmetric theory requires two complex Higgs doublets separately 
coupling to up-type and down-type fermions.  Out of the eight degrees 
of freedom, three result in the longitudinal components of the $W^{\pm}$ and $Z$ 
bosons.  The remaining five result in the following physical states: two neutral 
CP-even bosons, denoted by $h$ and $H$ in order of increasing mass; one neutral 
CP-odd boson ($A$); and two charged bosons ($H^{\pm}$).  
\par
The vacuum expectation values (vev's) of the Higgs fields give bosons and fermions 
their non-zero masses.  The ratio of vev's with up-type to down-type couplings is 
the supersymmetric parameter $\tan\beta$.  An important prediction of supersymmetry 
is that down-type couplings to Higgs bosons include a factor of $\tan\beta$.  Thus, 
down-type couplings are significantly enhanced at large $\tan\beta$, resulting in 
large production cross sections at the Tevatron.  In particular, for $\tan\beta = 50$ 
the cross section of a 100 GeV Higgs boson is about two orders of magnitude larger than 
in the SM.  In addition, $h$ ($H$) and $A$ are approximately mass-degenerate for $m_A$ 
less than (greater than) about 100 GeV, effectively doubling the production cross section.  
\par
Tevatron searches focus on the dominant decay modes of $b\bar{b}$ and $\tau\tau$ and 
are generally interpreted for a few different scenarios.  All scenarios maximize the 
Higgs-boson mass as a function of $\tan\beta$ in order to avoid the direct LEP limits.  
They either have non-vanishing (``$m_h^{max}$") or vanishing (``no-mixing") stop mixing, 
and probe both positive and negative values of the Higgs mixing parameter at the 
electroweak scale ($\mu$).

\subsection{Searches for $H \rightarrow \tau\tau$}
\label{sec:htautau}
\hspace*{0.18in}
CDF and D\O\ \cite{d0htautau} have searched for Higgs bosons decaying to tau 
pairs in final states where one tau decays to an electron ($\tau_e$) or muon 
($\tau_{\mu}$) and the other tau decays either hadronically ($\tau_h$) or to 
a different-flavor lepton.  Because most of the sensitivity comes from 
final states with $\tau_h$, these searches rely on identifying and measuring 
hadronic tau decays.
\par
CDF's $\tau_h$ identification starts from a reconstructed high-momentum ``seed" 
track.  Additional tracks within a narrow $\eta-\phi$ cone are associated with 
the tau decay, and the total momentum outside this cone (and within 
a larger cone) is required to be small.  The defined tau cone shrinks with 
increasing tau momentum, accounting for the larger boost.  The tau momentum 
is measured by combining the track momenta (typically from charged pions) 
with the electromagnetic calorimeter momenta (typically from neutral pions), 
with a correction for the expected charged-pion contribution to the electromagnetic 
momenta.
\par
D\O\ subdivides the $\tau_h$ decays into cases where a single charged track matches 
either energy deposited in the hadronic calorimeter (consistent with a charged pion) 
or energy in both the electromagnetic and hadronic calorimeters (consistent with 
both charged and neutral pions), or where there are three charged tracks with 
invariant mass less than the tau mass (1.7 GeV).  In each of these cases, D\O\ uses 
a neural network discriminant to separate tau decays from direct hadron production.  
In the case where there is significant measured electromagnetic energy, an additional 
neural network discriminates tau decays from direct electron production.  
\begin{wrapfigure}{r}{0.5\columnwidth}
\centerline{\includegraphics[width=0.45\columnwidth]{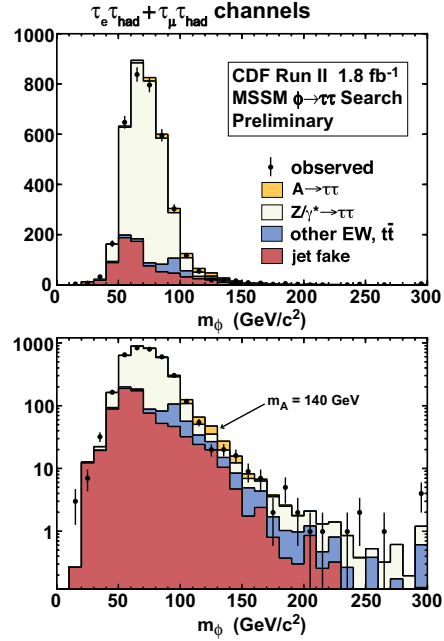}}
\caption{The CDF $\tau_h \tau_l$ invariant mass distribution on linear 
(top) and log (bottom) scales.}
\label{fig:cdfhtautau}
\vskip -0.1in
\end{wrapfigure}

\begin{figure}[!t]
\centerline{\includegraphics[width=0.5\columnwidth]{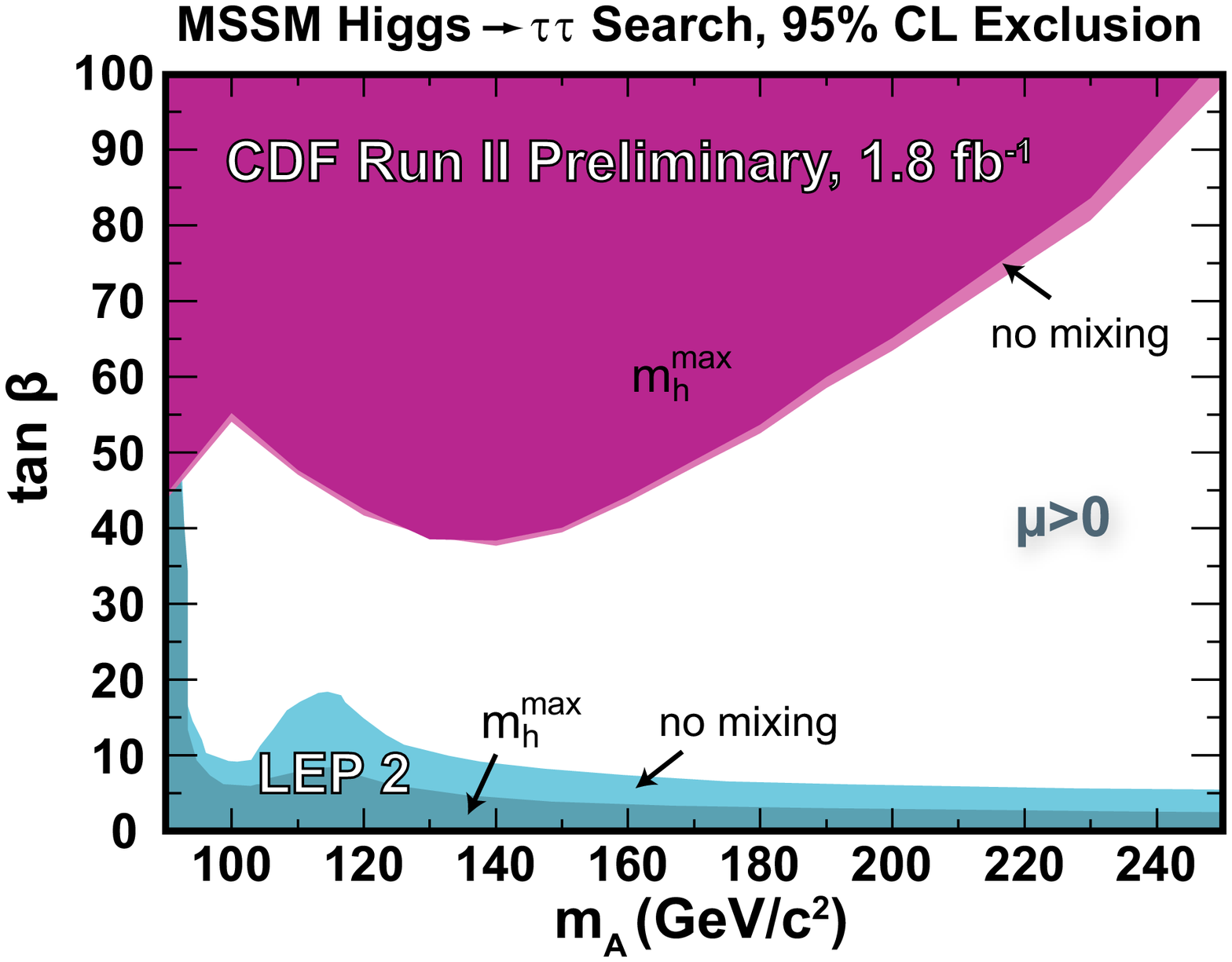}
   	    \includegraphics[width=0.44\columnwidth]{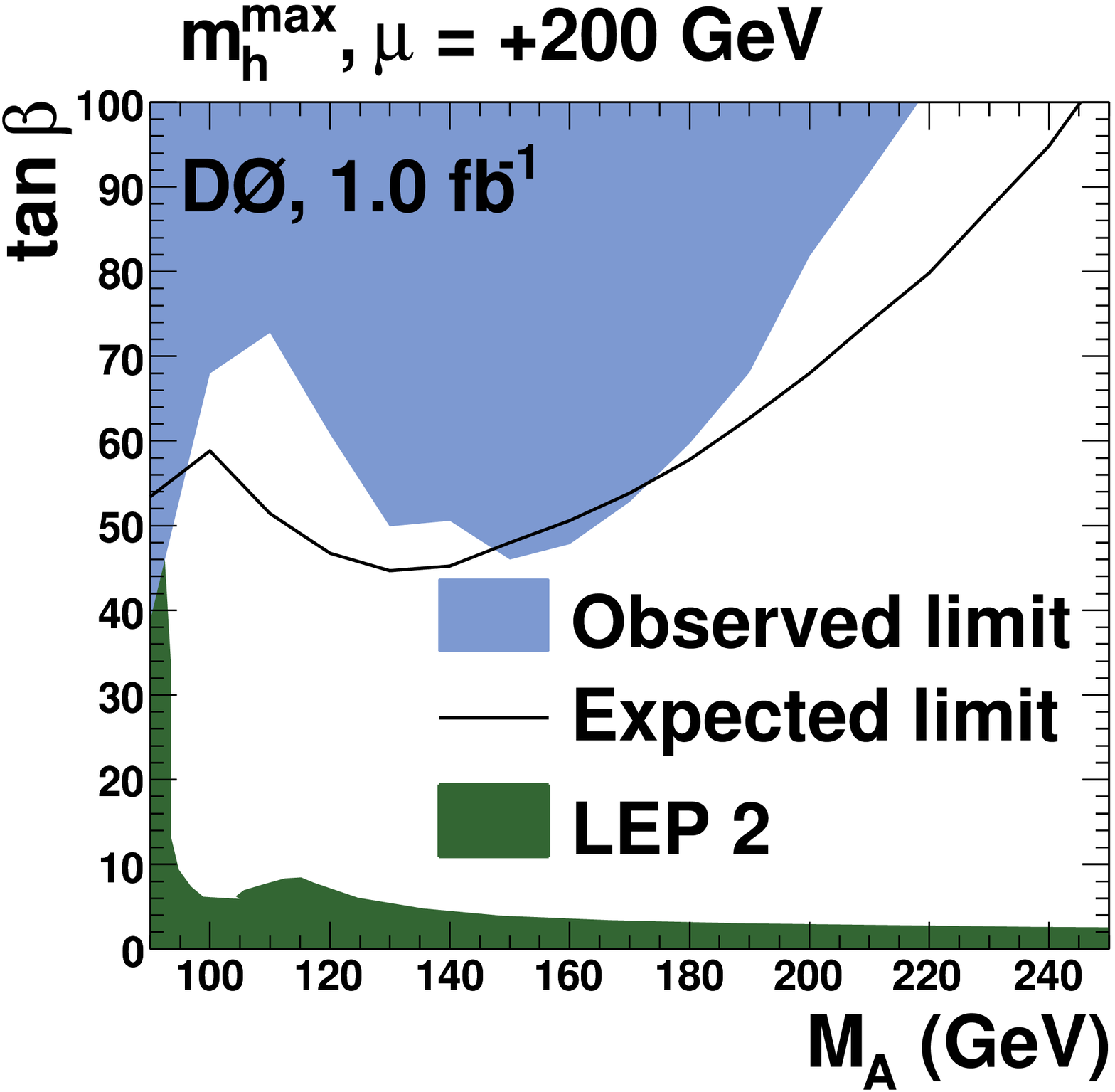}}
\caption{CDF (left) and D\O\ (right) limits in the $m_A - \tan\beta$ plane.  CDF examines 
the $m_h^{max}$ (dark pink) and no-mixing (light pink) scenarios with $\mu >0$ and $\mu < 0$
(not shown).  The $\mu < 0$ limits are similar.  D\O\ examines the $m_h^{max}$ and no-mixing 
(not shown) scenarios for $\mu > 0$.  The no-mixing limits are similar.}
\label{fig:cdfd0htautaulimits}
\end{figure}

Figure~\ref{fig:cdfhtautau} shows the CDF ditau invariant mass spectrum in events 
with one leptonic and one hadronic tau decay.  No excess consistent with Higgs 
production is observed in the CDF or D\O\ data, so limits are set in the 
$m_A-\tan \beta$ plane for the $m_h^{max}$ and no-mixing scenarios with $\mu > 0$ 
and $\mu < 0$ (Fig.~\ref{fig:cdfd0htautaulimits}).  

\subsection{Searches for $(b)bH \rightarrow (b)bbb$}
\label{sec:hbb}
\hspace*{0.18in}
The $p\bar{p} \rightarrow H \rightarrow b\bar{b}$ process is overwhelmed by direct 
$b\bar{b}$ production, so searches for this Higgs decay channel probe the processes 
$gb \rightarrow Hb \rightarrow bbb$ and $q\bar{q}/gg \rightarrow bbH \rightarrow bbbb$.  
CDF and D\O\ \cite{d0hbb} require at least three identified $b$ quarks in the final 
state, making $b$-quark identification a key component of the analysis.  In addition, 
$b$-jet energy resolution is important to resolve the Higgs mass peak from the 
background continuum, though for $\tan\beta \gtrsim 100$ the intrinsic width is larger 
than the detector resolution.
\par
CDF identifies $b$ quarks using the significance of the jet vertex displacement with 
respect to the collision vertex, combined with the invariant mass of the tracks used 
in the jet vertex reconstruction (the mass of $b$-quark vertices is higher than that of 
lighter quarks).  The background to the sample is predominantly heavy-flavor quark 
production, whose mass distributions are predicted using a combination of data and 
{\sc pythia} MC.  The results of a detailed measurement of azimuthal correlations in 
$b\bar{b}$ production are used to tune the gluon splitting contribution in the MC.
The mass spectrum of the two highest $E_T$ $b$-jets shows no deviation greater than 
$2\sigma$ between data and MC, so mass limits are set in the $m_A-\tan\beta$ plane 
for $\mu = -200$ GeV in the $m_h^{max}$ scenario.  The $\tan\beta$ limit increases 
from $\approx 80$ to $\approx 100$ for $m_A$ increasing from 110 GeV to 180 GeV.
\begin{wrapfigure}{r}{0.5\columnwidth}
\centerline{\includegraphics[width=0.45\columnwidth]{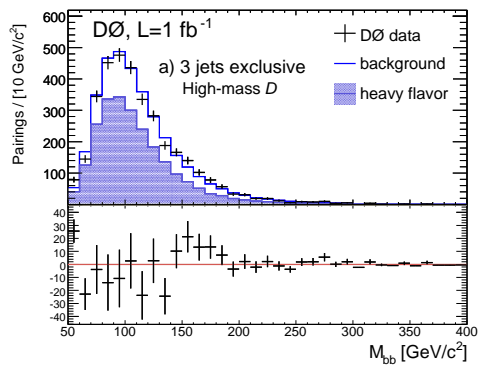}}
\caption{The D\O\ $m_{bb}$ distribution in the final state with exactly three jets, all 
identified as originating from a $b$ quark. }
\label{fig:d0hbb}
\vskip -0.1in
\end{wrapfigure}
\par
D\O\ uses a neural network discriminant to identify $b$ quarks.  The background is 
estimated separately for each jet multiplicity using the measured $b$-quark identification 
efficiency and control samples with and without identified $b$-jets.  Six variables are 
used in a likelihood to separate Higgs signal from background, and the resulting mass 
spectrum shows good agreement between data and predicted background, with a small excess 
around 180 GeV (Fig.~\ref{fig:d0hbb}).  The probability for this excess to correspond to a 
background fluctuation is 4.4\%.  Limits are set in the $m_A-\tan\beta$ plane for 
$\mu = \pm 200$ GeV for the $m_h^{max}$ (negative $\mu$ only) and no-mixing scenarios.  In 
the $m_H^{max}$ scenario, the limits range from $\tan\beta > 55$ for $m_A \approx 125$~GeV 
to $\tan\beta > 100$ for $m_A \approx 200$~GeV.

\section{Higgs Boson Searches in Left-Right Symmetric Models}
\label{sec:lrhiggs}
\hspace*{0.18in}
Left-right symmetric theories postulate the breaking of an SU(2) symmetry into 
the observed SU(2)$_{\rm L}$ weak symmetry and an unobserved SU(2)$_{\rm R}$ 
symmetry.  The SU(2)$_{\rm R}$ symmetry is broken by the vev of a weak 
SU(2)$_{\rm R}$ Higgs triplet.  The 
corresponding neutral Higgs boson couples only to neutrinos and bosons, and 
results in the see-saw mechanism producing small neutrino masses.  The triplet 
includes a doubly charged Higgs boson $H^{\pm\pm}$ that couples to like-sign 
lepton pairs.  In a supersymmetric model, these Higgs bosons can have masses of 
${\cal O}$(100 GeV), and thus be observable at the Tevatron collider.

\subsection{Searches for $H^{\pm\pm} \rightarrow l^{\pm}l^{\pm}$}
\label{sec:hpp}
\hspace*{0.18in}
\vskip -0.1in
\begin{wrapfigure}{r}{0.5\columnwidth}
\centerline{\includegraphics[width=0.45\columnwidth]{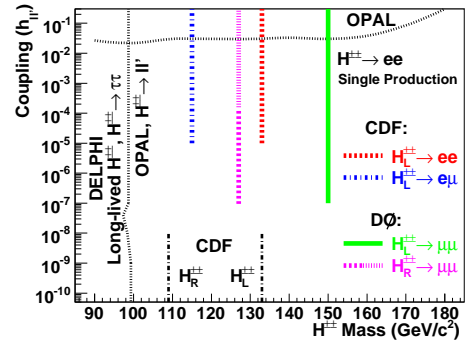}}
\caption{Limits from direct searches for $H^{\pm\pm}$ production, assuming 
exclusive decay to a given lepton pair.  Not shown are: CDF limits of
$m_{H^{\pm\pm}_L} > 114$~GeV and 112~GeV for decay to $e\tau$ and $\mu\tau$, 
respectively; and the H1 limit of $m_{H^{\pm\pm}_L} > 141$~GeV for a coupling 
$h_{e\mu} =  0.3$. }
\label{fig:hpp}
\vskip -0.1in
\end{wrapfigure}
Doubly charged Higgs bosons are predominantly produced in pairs at the Tevatron.
Events with four leptons in the final state have negligible background, and 
searches typically require fewer than four leptons to increase signal acceptance.
The D\O\ collaboration has recently performed a search in the three-muon final 
state \cite{d0hpp}, excluding $H^{\pm\pm}$ masses above 150 (127) GeV for Higgs 
bosons coupling to left-(right-) handed muons with 100\% branching ratio.  This 
result increases the earlier limit from D\O\ and complements CDF \cite{cdfhpp}, 
H1 \cite{h1hpp}, OPAL, and DELPHI searches in all $H^{\pm\pm}$ dilepton decay 
channels, and for long-lived $H^{\pm\pm}$ (Fig.~\ref{fig:hpp}).

\section{Summary}
\label{sec:summary}
\hspace*{0.18in}
There is strong motivation for additional Higgs bosons beyond that of the SM, 
and recent Tevatron searches have probed for their existence.  In addition to 
the supersymmetric and doubly-charged Higgs boson searches described here, D\O\ 
has searched for fermiophobic Higgs bosons.  With 2-10 times unsearched data 
available, more than one Higgs boson could be awaiting discovery at the 
Tevatron.

\begin{footnotesize}

\end{footnotesize}

\end{document}